# Generation of Hidden Optical-Polarization: Squeezing and Non-Classicality

Gyaneshwar K. Gupta[1*], Akhilesh Kumar[2], Ravi S. Singh[1#]

[1]Department of Physics, DDU Gorakhpur University, Gorakhpur-273009 (U.P.) – INDIA

[2]Department of Physics, Govt. P. G. College, Rishikesh-249201(U.K.) - INDIA

e-mail: [#]yesora27@gmail.com, [*]gyankg@gmail.com

**Abstract:** A monochromatic bi-modal coherent light, endowed with orthogonally polarized photons propagating collinearly, is allowed to impinge as pump-field in degenerate parametric amplification. Generation of Hidden Optical-Polarized State is seen by non-zero values of Index of Hidden Optical-Polarization. Squeezing in Hidden Optical-Polarized State is demonstrated by recognizing a Squeezing Function of which departure from unit-value ensures the squeezing. Furthermore, the non-classical feature is witnessed by the 'Degree of Hidden Optical-Polarization' having value 'greater than unity'.

**Keywords**: Optical Polarization, Hidden Optical-Polarization, Polarization Squeezing, Non-classicality,

## 1 Introduction

Polarization in Optics is an age-old concept discovered by Danish Mathematician E. Bartholinus and interpreted by Dutch Physicist C. Huygens through his conception of secondary spherical light wave while investigating birefringence in Quartz crystal [1]. Seminal contributions from Young, Arago, and Fresnel established the transversal nature (polarization structure) of the optical field, which is revealed by temporal evolution, at any spatial point, of electric field vector precessing an ellipse of non-random eccentricity and orientation in the transverse plane with respect to direction of propagation of parallel, uniform, monochromatic light beam [2]. Varying the non-random values of 'ratio of amplitudes' and 'difference in phases' of components of the optical field along orthogonal bases-modes by suitable combinations of rotator (polarizer) and/or phase-shifter (compensator), elliptically polarized light may be degenerated into states of linear and circular polarizations. The State of Optical-Polarization (SOP) in Classical Optics is characterized and quantified by Stokes parameters [3] of which experimental-measurements are well established. Although several techniques, namely, Jones matrix, Mueller matrix and Coherency matrix [4] are evolved for quantitative investigations of SOP, Stokes-parameters have acquired the inevitable positions since they find straightforward generalization to characterize optical field in quantum regime [5].

The study of polarization of light encompasses, along with two extreme cases: unpolarized light and polarized light, infinitely many SOP which are neither polarized nor unpolarized. Until 1970s severe confusions prevailed pertaining to the rigorous definition of unpolarized light. In 1971 Prakash and Chandra [6], and independently Agarwal [7], discovered the structure of density operator for unpolarized light by demanding invariance of complete statistical properties, which depends upon correlations between higher-order moments of optical-field amplitudes, about the direction of propagation. Furthermore, the authors [6] categorically stressed that Stokes parameters are insufficient in characterizing the SOP, especially, where correlations between optical field-amplitudes of higher-order are crucial. Various studies on the state of unpolarized light [8-9] have been performed bringing in new insights about its quantum nature. Mehta and Sharma [10] defined polarized light stringently by requiring disappearance of light signal in at least one transverse orthogonal mode, although the treatment doesn't provide a prescription for testing whether an arbitrary quantum state of light is polarized or unpolarized.

Moreover, the studies on optical-polarization of quantum fields witnessed conflicting yet complementary approaches: Computable-measures and Operational-measures. Numerous computable-measures [11-12], based on the abstract notion of distance between quantum state in question to that of unpolarized optical field, are introduced giving variant expressions for degree of polarization which, in turn, assess the SOP. The 'distance based approach' assigns not only different values of degree of polarization for the same quantum state in question but also its correspondence to classical-description of the optical-polarization couldn't be drawn. On the other hand, the operational-measure is carried out by experimentally measured-values of Stokes parameters for quantum fields despite of the consensual fact that Stokes parameters are inadequate [13-15] and absurd conclusions may appear by usage of them for the characterization of SOP in quantum domain vis-a-vis to classical regime [16]. Nevertheless Stokes parameters furnish informations obtained by instantaneous correlations between optical field modes at any spatial point. Wolf [17] has unified the theory of coherence and polarization which, being only a second-order theory, can explain arbitrary alteration in SOP of the light beam undergoing propagation through space. Quite recently Klimov et al. [18] formulated, and verified experimentally, an innovative criterion for degree of polarization in terms of minimal fluctuations in Stokes parameters on Poincare sphere and advocate that the Stokes parameters are basic tools for study of optical-polarization. These authors [18] speculate that the laid-down criterion solves almost all imprecise knowledge of optical-polarization of quantum fields where second-order correlations in Stokes parameters (forth-order correlations in field-amplitudes) play indispensable roles. Radical observations about higher-order correlations in Stokes parameters might point out huge conceptual lackings. Unless higher-order Stokes parameters behave synonymously like Glauber correlation functions [19], the characterization of optical-polarization must be carried out by alternative viable rational formulation [20]. Prakash and Singh [21] adopts a fundamental departure, intrinsically in adherence with the classical description of optical-polarization, from traditional Stokes parameters' approach and worked out an optical-polarization operator expressed as the product of Bosonic inverse-annihilation operator [22] and usual annihilation operator along the transverse orthogonal modes in a basis of description. This optical-polarization operator not only sets up a criterion for deciphering whether a light in any arbitrary quantum state is perfectly polarized by satisfying the modified eigen-value equation but also it picks up the 'characteristic-parameters' for perfect optical-polarization states, i. e. the non-random values of the 'ratio of real amplitudes' and 'difference in phases' in a basis of description [23].

Klyshko [24] has coined the two inequivalent and dissimilar terms: 'hidden polarization' and higher-order polarization (Malus laws) for optical fields possessing non-zero anisotropic values of higher-order intensity correlations' functions. Bjork et al. [12], while surveying intensively various impeccable proposals for quantum degrees of polarization, opined that 'it would be better to say that such states (bi-modal Fock states in Klyshko's analysis) preserve 'higher-order polarization' rather 'hidden polarization'. Notably, Singh and Prakash [25] have generalized the usual theory of optical-polarization to study the polarization properties of light in macroscopic superposition of coherent states such as Schrodinger cat states, Even- and Odd-coherent states etc.. The definition [Eq.12 in Ref.25] of higher-order optical polarization can be employed to extract the polarization structure envisaged for bi-modal Fock states in Klyshko's theory.

Heisenberg uncertainty relations among canonically conjugate dynamical observables provide the minimal bound for dispersion (noise). The concept of Squeezing [26] is introduced to seek the possibility of reducing noise in field-observables below those possessed by light at vacuum or at coherent states. Squeezing in noise of one observable induces increased noise (anti-squeezing) in other observable indicating more fuzziness in its measurement. The studies on squeezing and non-classicality [27] received considerable interests which pave the way for generation and application of non-classical sources of light in quantum metrology for beating the limit set by Heisenberg uncertainty principle [28]. Stokes-operators obey SU (2) Lie algebra and, therefore, simultaneous measurements with extreme precisions are impossible which, in turn, render the inaccuracy in characterization of optical-polarization by them. The squeezing of one or more Stokes parameters below than those in vacuum or coherent state of the optical field is understood as the polarization-squeezed state. To enhance the sensitivity of polarization-interferometer Grangier et al. [29] generated polarization-squeezed beam in the optical parametric process. Korolkova et al. [30], theoretically, predicts the generation of polarization-squeezed beam by interfering two orthogonally polarized intense quadrature-squeezed beams [31-32]. Bowen et al. [33] implemented, experimentally, the Korolkova's scheme for generation of polarization-squeezed beam of which pairs combine at the beam splitter to produce continuous-variable polarization entanglement.

Prakash et al. [34] introduced a single-mode monochromatic optical field of which 'characteristics parameters' are the non-random values of 'ratio of real amplitudes' and 'sum of phases' rather than those of 'ratio of real amplitudes' and 'difference in phases' (characteristics parameters for polarized light). All Stokes parameters for such an optical-field are seen to have vanishing-values except one which equals intensity of the optical-field. Evidently, one may arrive at absurd conclusions and ascribe the state of such an optical-field as unpolarized one which is not truth. The authors [34] have used the generic term 'Hidden Optical-Polarization state (HOPS)' for such light of which polarization nature can't be understood by Stokes parameters [35]. The present paper is intended to study the generation of HOPS and investigates the possibility of squeezing and non-classicality therein, when bi-modal coherent light evolves in degenerate optical parametric-amplification. The paper is organized as follows: Section 2 describe the precise formulations for definitions of Indices of optical-polarization and Hidden optical-polarization. In section 3, Inadequacy of Stokes parameters for HOPS are highlighted which induces to introduce Hermitian operators for its operational-characterization. Section 4 describes the generation of HOPS. Section 5 deals with squeezing in HOPS analogous to squeezing in Stokes parameters, by identifying a squeezing function defined to satisfy inequality inspired by Heisenberg uncertainty principle. Section 6 inquires an alternative non-classical signature, obtained by acquiring greater than unit-value for degree of Hidden optical-polarization. Finally, some conclusions are drawn.

## 2 Indices of Optical-Polarization and Hidden Optical-Polarization

A plane monochromatic unpolarized uniform optical field propagating along z-direction can, in general, be described by vector potential (analytical signal), $\mathcal{A}$,

$$\mathcal{A} = (\hat{\mathbf{e}}_x \underline{A}_x + \hat{\mathbf{e}}_y \underline{A}_y) e^{-i\psi}, \quad \underline{A}_{x,y} = A_{0x,0y}\, e^{i\phi_{x,y}}, \qquad (1)$$

where $\psi = \omega t - kz$, $\underline{A}_{x,y}$ are classical complex amplitudes (CAs), $A_{0x,0y}$ are real amplitudes undergoing, in general, random spatio-temporal variation with angular frequency, $\omega$, $\phi_{x,y}$ are phase parameters which may take equally probable random values, $0 \leq \phi_{x,y} < 2\pi$ in linear polarization-basis $(\hat{\mathbf{e}}_x, \hat{\mathbf{e}}_y)$, $\mathbf{k}$ $(= k\hat{\mathbf{e}}_z)$ is propagation vector of magnitude k, and $\hat{\mathbf{e}}_{x,y,z}$ are unit vectors along respective x-, y-, z-axes forming right handed triad. Vividly, optical field, Eq.(1) is representative of bi-modal unpolarized optical field because it needs two random CAs, $\underline{A}_{x,y}$ in transverse orthogonal basis-modes $(\hat{\mathbf{e}}_{x,y}\ \bar{k})$ for its complete statistical-characterization. Optical field may be said to be polarized only when the ratio of CAs must keep non-random values, i.e.

$$p = \underline{A}_y / \underline{A}_x, \qquad (2)$$

where p is a non-random complex parameter defining 'index of polarization' (IOP) in the linear-polarization basis $(\hat{\mathbf{e}}_x, \hat{\mathbf{e}}_y)$ [21]. Evidently, polarized optical field is a mono-modal optical field as only one random CA suffices for its complete statistical description (other orthogonal CA is specified by p). If one introduces new parameters $A_0$ (real random amplitude), $\chi_0$ (polar angle), $\Delta_0$ (azimuth angle), $\phi$ (random phase) on Poincare sphere, satisfying inequalities $0 \leq A_0$, $0 \leq \chi_0 \leq \pi$, $-\pi < \Delta_0 \leq \pi$, $0 \leq \phi < 2\pi$, respectively and preserving transforming equations in terms of old parameters, $A_0 = (A_{0x}^2 +$

$A_{0y}^2)^{1/2}$, $\chi_0 = 2 \tan^{-1}(A_{0y}/A_{0x})$ and $\Delta_0 = \phi_y - \phi_x$; $\phi = (\phi_x + \phi_y)/2$, the analytic signal, $\mathcal{A}$, Eq. (1), finds an instructive compact form

$$\mathcal{A} = \hat{\boldsymbol{\varepsilon}}_0 \mathcal{A}; \; \mathcal{A} = \underline{A} e^{-i\psi}; \; \underline{A} = A_0 \, e^{i\phi},$$
$$\hat{\boldsymbol{\varepsilon}}_0 = \hat{\mathbf{e}}_x \cos\frac{\chi_0}{2} e^{-\Delta_0/2} + \hat{\mathbf{e}}_y \sin\frac{\chi_0}{2} e^{-\Delta_0/2}. \qquad (3)$$

The Eq. (3) may be interpreted as the mono-modal polarized optical field, statistically described by single CA, $\underline{A}$ and polarized in the fixed direction, $\hat{\boldsymbol{\varepsilon}}_0$ determined by non-random angle parameters $\chi_0$ and $\Delta_0$ in the Poincare sphere and, thus, specifying the mode, $(\hat{\boldsymbol{\varepsilon}}_0, \mathbf{k})$. The complex vector $\hat{\boldsymbol{\varepsilon}}_0$ is a unit vector ($\hat{\boldsymbol{\varepsilon}}_0^* \cdot \hat{\boldsymbol{\varepsilon}}_0 = 1$, asterisk, * denotes complex conjugate) giving expression of IOP, $p = \tan\frac{\chi_0}{2} e^{i\Delta_0}$. Obviously, the SOP is specified by the non-random values of p, which, in turn, is fixed by non-random values of $\chi_0$ and $\Delta_0$ defining a point $(\hat{\boldsymbol{\varepsilon}})$, in the unit Poincare sphere, similar to Stokes parameters. All typical parameters in ellipsometry such as major axis, minor axis and orientation angles of the polarization-ellipse can be determined if p of optical field and one CA are specified. In elliptic-polarization basis, $(\hat{\boldsymbol{\varepsilon}}, \hat{\boldsymbol{\varepsilon}}_\perp)$ [36] such polarized light (Eq.3) retain IOP, $p_{(\hat{\boldsymbol{\varepsilon}},\hat{\boldsymbol{\varepsilon}}_\perp)}$, another non-random parameter showing clear dependence on $\hat{\boldsymbol{\varepsilon}}_0$, as

$$p_{(\hat{\boldsymbol{\varepsilon}},\hat{\boldsymbol{\varepsilon}}_\perp)} = \underline{A}_{\hat{\boldsymbol{\varepsilon}}_\perp}/\underline{A}_{\hat{\boldsymbol{\varepsilon}}} = \frac{\hat{\boldsymbol{\varepsilon}}_\perp^* \cdot \hat{\boldsymbol{\varepsilon}}_0}{\hat{\boldsymbol{\varepsilon}}^* \cdot \hat{\boldsymbol{\varepsilon}}_0}, \qquad (4)$$

where $\hat{\boldsymbol{\varepsilon}}$ and $\hat{\boldsymbol{\varepsilon}}_\perp$ are orthogonal complex unit vectors ($\hat{\boldsymbol{\varepsilon}}^* \cdot \hat{\boldsymbol{\varepsilon}} = \hat{\boldsymbol{\varepsilon}}_\perp^* \cdot \hat{\boldsymbol{\varepsilon}}_\perp = 1$, $\hat{\boldsymbol{\varepsilon}}_\perp^* \cdot \hat{\boldsymbol{\varepsilon}} = 0$). Formula, Eq.(4) caters interchangeability of IOP's between bases of descriptions. Parametrizing CAs $\underline{A}_{\hat{\boldsymbol{\varepsilon}}}$ and $\underline{A}_{\hat{\boldsymbol{\varepsilon}}_\perp}$ by $\underline{A}_{\hat{\boldsymbol{\varepsilon}}} = A_{0\hat{\boldsymbol{\varepsilon}}} \exp(i\phi_{\hat{\boldsymbol{\varepsilon}}})$, $\underline{A}_{\hat{\boldsymbol{\varepsilon}}_\perp} = A_{0\hat{\boldsymbol{\varepsilon}}_\perp} \exp(i\phi_{\hat{\boldsymbol{\varepsilon}}_\perp})$, with real-amplitudes $(A_{0\hat{\boldsymbol{\varepsilon}}}, A_{0\hat{\boldsymbol{\varepsilon}}_\perp})$ and phase-parameters $(\phi_{\hat{\boldsymbol{\varepsilon}}}, \phi_{\hat{\boldsymbol{\varepsilon}}_\perp})$, the Eq. (4) reveals that such a polarized beam maintain non-random values of (i) 'ratio of real amplitudes', $A_{0\hat{\boldsymbol{\varepsilon}}_\perp}/A_{0\hat{\boldsymbol{\varepsilon}}}$, and (ii) 'difference in phase', $(\phi_{\hat{\boldsymbol{\varepsilon}}_\perp} - \phi_{\hat{\boldsymbol{\varepsilon}}})$, in elliptic-polarization basis, $(\hat{\boldsymbol{\varepsilon}}, \hat{\boldsymbol{\varepsilon}}_\perp)$. The preceding discussions highlight the fact that the state of polarized optical field (Eq.3) may be described by considering any basis of description.

HOPS [34], being a mono-modal optical field, has non-random values of 'ratio of real amplitudes' and 'sum of phases' contrary to single-mode ordinary polarized light where non-random values of 'ratio of real amplitudes' and 'difference in phases' of components of the optical field along transverse orthogonal bases-modes served as characteristic polarization-parameters. The characteristic parameters for HOPS can be dovetailed to definite 'Index of Hidden Optical-Polarization' (IHOP) in elliptic-polarization basis $(\hat{\boldsymbol{\varepsilon}}, \hat{\boldsymbol{\varepsilon}}_\perp)$ as

$$p_{h(\hat{\boldsymbol{\varepsilon}}, \hat{\boldsymbol{\varepsilon}}_\perp)} = \underline{A}_{\hat{\boldsymbol{\varepsilon}}_\perp}/\underline{A}^*_{\hat{\boldsymbol{\varepsilon}}} = \tan\frac{\chi_h}{2} e^{i\Delta_h}, \qquad (5)$$

where $\chi_h$ and $\Delta_h$ are non-random angle parameters ($0 \leq \chi_h \leq \pi$ and $-\pi < \Delta_h \leq \pi$). The condition may be made more comprehensible if one expresses real amplitudes $(A_{0\hat{\varepsilon}}, A_{0\hat{\varepsilon}_\perp})$ and phase parameters $(\varphi_{\hat{\varepsilon}}, \varphi_{\hat{\varepsilon}_\perp})$ as

$$A_{0\hat{\varepsilon}} = A_0 \cos \chi_h/2,\ A_{0\hat{\varepsilon}_\perp} = A_0 \sin \chi_h/2;\ \varphi_{\hat{\varepsilon}} = \zeta + \Delta_h/2,\ \varphi_{\hat{\varepsilon}_\perp} = -\zeta + \Delta_h/2, \qquad (6)$$

where $A_0$ and $\zeta$ are random parameters ($0 \leq A_0$, $0 \leq \zeta < 2\pi$) satisfying $A_0 = (A^2_{0\hat{\varepsilon}_\perp} + A^2_{0\hat{\varepsilon}})^{1/2}$, $\chi_h = 2 \tan^{-1}(A_{0\hat{\varepsilon}_\perp}/A_{0\hat{\varepsilon}})$, and $2\zeta = \varphi_{\hat{\varepsilon}} - \varphi_{\hat{\varepsilon}_\perp}$, $\Delta_h = \varphi_{\hat{\varepsilon}_\perp} + \varphi_{\hat{\varepsilon}}$. The analytic signal of vector potential of HOPS can, explicitly, be written in elliptic-polarization basis, $(\hat{\varepsilon}, \hat{\varepsilon}_\perp)$ as

$$\mathcal{A} = [\hat{\varepsilon} \cos \frac{\chi_h}{2} A_0 e^{i\zeta} e^{i\Delta_h/2} + \hat{\varepsilon}_\perp \sin \chi_h/2\ A_0 e^{-i\zeta} e^{i\Delta_h/2}] e^{-i\psi}. \qquad (7)$$

Eq.(7) expresses analytic signal, $\mathcal{A}$ for the mono-modal optical field in which 'difference in phases', $\zeta$ is random parameter and sum of the phases, $\Delta_h$ is non-random, a distinctive unusual property of HOPS, but its complete statistical properties are described by one random CA ($A_0\ e^{i\zeta}$) or a random real amplitudes, $A_0$ and a random phase parameter, $\zeta$.

In Quantum Optics the optical field, Eq. (1) is described by operatic-version of vector potential operator,

$$\widehat{\mathcal{A}} = \left(\frac{2\pi}{\omega V}\right)^{1/2} [(\hat{\mathbf{e}}_x \hat{a}_x + \hat{\mathbf{e}}_y \hat{a}_y) e^{-i\psi} + \text{h.c.}],$$

$$= \left(\frac{2\pi}{\omega V}\right)^{1/2} [(\hat{\varepsilon} \hat{a}_{\hat{\varepsilon}} + \hat{\varepsilon}_\perp \hat{a}_{\hat{\varepsilon}_\perp}) e^{-i\psi} + \text{h.c.}], \qquad (8)$$

in linear-polarization basis $(\hat{\mathbf{e}}_x, \hat{\mathbf{e}}_y)$ or in elliptic-polarization basis $(\hat{\varepsilon}, \hat{\varepsilon}_\perp)$, respectively, where $\omega$ is angular frequency of the optical field and V is the quantization volume of the cavity, h.c. stands for Hermitian conjugate (denoted by dagger, †). Using orthonormal properties of $\hat{\varepsilon}$ $(= \varepsilon_x \hat{\mathbf{e}}_x + \varepsilon_y \hat{\mathbf{e}}_y)$ and $\hat{\varepsilon}_\perp (= \varepsilon_{\perp x} \hat{\mathbf{e}}_x + \varepsilon_{\perp y} \hat{\mathbf{e}}_y)$ the annihilation operators $\hat{a}_{\hat{\varepsilon}}$ ($\hat{a}_{\hat{\varepsilon}_\perp}$) are seen to be related with those in linear-polarization basis $(\hat{\mathbf{e}}_x, \hat{\mathbf{e}}_y)$ by,

$$\hat{a}_{\hat{\varepsilon}} = \varepsilon_x^* \hat{a}_x + \varepsilon_y^* \hat{a}_y,\ \hat{a}_{\hat{\varepsilon}_\perp} = \varepsilon_{\perp x}^* \hat{a}_x + \varepsilon_{\perp y}^* \hat{a}_y, \qquad (9)$$

satisfying usual standard Bosonic-commutation relations.

While enunciating quantum theory of optical coherence Glauber [19] defined correlation functions of arbitrary order which may be utilized to display correlation properties between CAs of bi-modal optical-field, Eq.(8) as,

$$\Gamma^{(m_x, m_y, n_x, n_y)} = \text{Tr}[\rho(0) \hat{\mathcal{A}}_x^{(-)m_x} \hat{\mathcal{A}}_y^{(-)m_y} \hat{\mathcal{A}}_x^{(+)n_x} \hat{\mathcal{A}}_y^{(+)n_y}] \qquad (10)$$

where $\rho(0)$ is density operator specifying the quantum state of optical field, $m_{x,y}$ and $n_{x,y}$ are arbitrary non-negative integers. Setting the condition on quantized complex amplitudes (annihilation operators), $\hat{a}_{x,y}$ by

$$\hat{a}_y(t)\rho(0) = p\hat{a}_x(t)\rho(0), \tag{11}$$

where p is IOP, and inserting Eq.(11) into (10) one obtains correlation functions

$$\Gamma^{(m_x,m_y,n_x,n_y)} = p^{*m_y}p^{m_y}\Gamma^{(M,\ 0,\ N,\ 0)}, \tag{12}$$

where M= $m_x$ + $m_y$ and N= $n_x$ + $n_y$. Clearly, above correlation functions, Eq.(12) for optical field imbibing the criterion Eq.(11) is determined by IOP, p and one of quantized complex amplitudes $\hat{a}_x(t)$ revealing once again that polarized light is mono-modal optical field. It is, therefore, the criterion Eq.(11) may be viewed as quantum analogue of classical criterion $\underline{A}_y = p\,\underline{A}_x$ for polarized optical field in linear polarization basis $(\hat{\mathbf{e}}_x, \hat{\mathbf{e}}_y)$. Multiplying Eq.(11) by inverse annihilation operator $\hat{a}_x^{-1}(t)$ from the left one obtains the criterion, $\hat{P}\rho(0) = p\,(1-\hat{V}_x)\,\rho(0)$, where $\hat{V}_x$ is the vacuum projection operator for x-linearly polarized virtual photons, satisfied by the optical polarization operator $\hat{P} \equiv \hat{a}_x^{-1}(t)\,\hat{a}_y(t)$ [21]. Similarly, having employed the criterion,

$$\hat{a}_y(t)\rho(0) = p_h e^{-2i\omega t}\rho(0)\hat{a}^\dagger_x(t), \tag{13}$$

where $p_h$ is IHOP in linear polarization basis, $(\hat{\mathbf{e}}_x, \hat{\mathbf{e}}_y)$, and substituting Eq.(13) into Eq.(10), we obtain correlation functions for single-mode HOPS as,

$$\Gamma^{(m_x,m_y,n_x,n_y)} = p_h^{*m_y}p_h^{n_y}\Gamma^{(M,\ 0,\ N,\ 0)} \tag{14}$$

Eq.(13) must, following similar cogent reasoning as above, be recognized as the quantum counterpart of the classical equation, $\underline{A}_y = p_h\,\underline{A}_x^{*}$ for hidden optical-polarized field in linear polarization basis $(\hat{\mathbf{e}}_x, \hat{\mathbf{e}}_y)$ [cf. Eq.(5)]. Moreover, the criterion Eq.(13) for Hidden optical-polarized field may be applied to compute the IHOP ($p_h$) for quantum fields generated in degenerate parametric amplification (see below Eq.(22)).

## 3 Hidden Optical-Polarization Parameters

Traditionally, Stokes parameters $s_{0,1,2,3}$ : $s_{0,1} = <|\underline{A}_y|^2 \pm |\underline{A}_x|^2>$, $s_2 + i\,s_3 = 2<\underline{A}_y^* \underline{A}_x>$, which took the roles of Hermitian operators (Stokes operators), $\hat{S}_{0,1,2,3}$: $\hat{S}_{0,1} = \hat{a}^\dagger_y(t)\hat{a}_y(t) \pm \hat{a}^\dagger_x(t)\hat{a}_x(t)$ and $\hat{S}_2 + i\hat{S}_3 = 2\hat{a}^\dagger_y(t)\hat{a}_x(t)$ giving quantum average values $s_{0,1,2,3}$ defined by $s_{0,1} = <\hat{S}_{0,1}> = \text{Tr}\,[\rho\,(0)\,\{\hat{a}^\dagger_y(t)\hat{a}_y(t) \pm \hat{a}^\dagger_x(t)\hat{a}_x(t)\,\}]$, $s_2 + i\,s_3 = <\hat{S}_2 + i\hat{S}_3> = 2\text{Tr}\,[\rho\,(0)\,\hat{a}^\dagger_y(t)\hat{a}_x(t)]$ for the optical field $\rho(0)$, characterize the SOP in Classical (Quantum) optics respectively, where < > provides ensemble (quantum) averages, Tr stands for trace of parenthesized quantities and $\underline{A}_{x,y}$ ($\hat{a}_{x,y}(t)$) gives classical (quantized) CAs of optical field in linear-polarization basis ($\hat{e}_x, \hat{e}_y$). If Stokes parameters were applied to ascertain the SOP of field, Eq(7), one would, paradoxically, conclude to assign it's state as unpolarized light. Considering non-random vanishing angle parameters on Poincare sphere as $\chi_h = 0 = \Delta_h$ and the elliptical-polarization basis ($\hat{\varepsilon}, \hat{\varepsilon}_\perp$) as the linear-polarization basis ($\hat{e}_x, \hat{e}_y$), the Stokes parameters defined by above expressions, noting the fact that random variables φ is equally likely between 0 to 2π, may be seen to have values, $s_0 = A_0^2$ and $s_1 = s_2 = s_3 = 0$. Evidently, at first glance, these values of Stokes parameters demonstrate that the light (Eq.7) is in unpolarized state which is not the tangible fact because light is actually in HOPS.

The moot point is that to decipher the polarization structure of HOPS usual Stokes parameters' approach must be amended. IHOP, Eq(5) may be one of the computable measures which predicts the requisite characteristics parameters: 'ratio of real amplitudes' and 'sum of phases' in any basis of description, which attributes a direction in the Poincare sphere. The operational-measure for HOPS may be obtained by the following Hidden Optical–Polarization parameters (cf. Stokes parameters), $h_{0,1,2,3}$ defined, in Classical optics, as

$$h_{0,1} = <|\underline{A}_y|^2 \pm |\underline{A}_x|^2>; \quad h_2 + i\, h_3 = 2<\underline{A}_y\, \underline{A}_x> \tag{15}$$

which may be generalized into operators, $\widehat{H}_{0,1,2,3}$ by

$$\widehat{H}_{0,1} = \hat{a}^\dagger_y(t)\hat{a}_y(t) \pm \hat{a}^\dagger_x(t)\hat{a}_x(t); \quad \widehat{H}_2 + i\, \widehat{H}_3 = 2\, e^{2i\omega t}\, \hat{a}_y(t)\hat{a}_x(t), \tag{16}$$

providing quantum average values $h_{0,1,2,3}$, by expressions,

$$h_{0,1} = <\widehat{H}_{0,1}> = \mathrm{Tr}[\rho(0)\{\hat{a}^\dagger_y(t)\hat{a}_y(t) \pm \hat{a}^\dagger_x(t)\hat{a}_x(t)\}];$$

$$h_2 + i\, h_3 = <\widehat{H}_2 + i\, \widehat{H}_3> = 2\, e^{2i\omega t}\, \mathrm{Tr}\,[\rho(0)\, \hat{a}_y(t)\, \hat{a}_x(t)]. \tag{17}$$

Following the proposals for measurements of Stokes parameters by Korolkova et al.[30], Singh and Gupta [37] formally proposed experimental arrangement for measuring the Hidden Optical-Polarization parameters. The Hidden Optical-Polarization Operators (Eq.(16)) can be demonstrated to share commutation relations,

$[\hat{H}_1, \hat{H}_0] = [\hat{H}_1, \hat{H}_2] = [\hat{H}_1, \hat{H}_3] = 0, [\hat{H}_0, \hat{H}_2] = 2i\hat{H}_3, [\hat{H}_0, \hat{H}_3] = 2i\hat{H}_2, [\hat{H}_2, \hat{H}_3] = 2i(\mathbb{1}+\hat{H}_0)$, (18)

and relationship $\hat{\mathbf{H}}^2 - \hat{H}_0^2 = 2(\mathbb{1} + \hat{H}_0)$, where $\hat{\mathbf{H}}(\hat{H}_1, \hat{H}_2, \hat{H}_3)$ is Hidden optical-polarization vector and, $\mathbb{1}$ is identity operator. Comparing Eq(18) with the standard SU (2) Lie algebraic equations for Stokes operators, namely, $[\hat{S}_0, \hat{S}_1] = [\hat{S}_0, \hat{S}_2] = [\hat{S}_0, \hat{S}_3] = 0; [\hat{S}_1, \hat{S}_2]=2i\hat{S}_3, [\hat{S}_2, \hat{S}_3] = 2i\hat{S}_1, [\hat{S}_3, \hat{S}_1] =2i\hat{S}_2$, one should take cognizance that hidden-polarization operator $\hat{H}_1$ commutes with all others $\hat{H}_{0,2,3}$ contrary to $\hat{S}_0$ bearing similar property. Thus, non-commutability of Hidden optical-polarization parameters ensures that their simultaneous measurements with unbounded precisions are impossible, which may be discriminated to introduce the concept of squeezing in HOPS. Heisenberg uncertainty principle can be invoked to give bounds for dispersion (noise) products for hidden optical-polarization parameters as,

$$\langle (\Delta \hat{H}_0)^2 \rangle \langle (\Delta \hat{H}_2)^2 \rangle \geq |\langle \hat{H}_3 \rangle|^2, \langle (\Delta \hat{H}_2)^2 \rangle \langle (\Delta \hat{H}_3)^2 \rangle \geq |\langle \mathbb{1} + \hat{H}_0 \rangle|^2,$$

$$\langle (\Delta \hat{H}_3)^2 \rangle \langle (\Delta \hat{H}_0)^2 \rangle \geq |\langle \hat{H}_2 \rangle|^2, \quad (19)$$

where $\langle (\Delta \hat{H}_j)^2 \rangle = \langle (\hat{H}_j)^2 \rangle - (\langle \hat{H}_j \rangle)^2$ is a shorthand notation for the dispersion (noise) of the parameter $\hat{H}_j$ (j = 0, 1, 2, 3), whose square root provide the uncertainty in the measurements of hidden optical-polarization parameters.

**4 Generation of Hidden Optical-Polarized States**

A bi-modal coherent optical field, endowed with photons having orthogonal polarization and propagating collinearly, is allowed to pump non-linear anisotropic crystal in degenerate parametric amplification. Following Glauber and Mollow [38] the Hamiltonian of the process may be considered as,

$$H = \omega[\hat{a}^\dagger_x(t)\hat{a}_x(t) + \hat{a}^\dagger_y(t)\hat{a}_y(t)] + k[\hat{a}_x(t)\hat{a}_y(t)e^{2i\omega t} + h.c.], \quad (20)$$

where k is real coupling constant proportional to second ordered nonlinear susceptibility ($\chi^{(2)}$) of the anisotropic crystal. The exact solutions of equations of motion, $i\dot{\hat{a}}_{x,y} = [\hat{a}_{x,y}, \hat{H}]$ of quantized complex amplitudes are,

$$\hat{a}_x(t) = e^{-i\omega t}[C(2)\hat{a}_x - iS(2)\hat{a}^\dagger_y]; \quad \hat{a}_y(t) = e^{-i\omega t}[C(2)\hat{a}_y - iS(2)\hat{a}^\dagger_x], \quad (21)$$

where over dot (.) represents time-variation and usage of natural convention c = ℏ = 1 is adopted, $\hat{a}_{x,y} \equiv \hat{a}_{x,y}(t = 0)$, C(2l) and S(2l) are hyperbolic time-varying functions defined by C(2l) ≡ Cosh2lkt and S(2) ≡ Sinh2lkt; for positive integer l. Taking bi-modal coherent light in pure state, $\rho(0) = |\alpha_x, \alpha_y\rangle\langle\alpha_x, \alpha_y|$, and inserting Eqs.(21) in Eq.(13) one obtains, after simple algebraic manipulations, the IHOP at an arbitrary interaction time t,

$$p_h(t) = \frac{p_h - i\, T(2)}{1 + i\, p_h\, T(2)}, \quad (22)$$

where $p_h \equiv p_h(0) = \alpha_y / \alpha_x^*$, $T(2) \equiv \tanh 2kt$, of which non-vanishing values at an arbitrary interaction time assures the generation of HOPS. Since IHOP, $p_h(t)$ being a complex non-random parameter, the absolute magnitude, $|p_h(t)|$ and the phase, $\arg(p_h(t))$, i.e. the 'ratio of real amplitudes' and 'sum of phases' in linear polarization basis $(\hat{e}_x, \hat{e}_y)$ deciphers the 'characteristics parameters' for generated HOPS.

## 5 Squeezing in Hidden Optical-Polarized states of light

Light is said to be in squeezed state if the dispersion (noise) of one or more of the hidden optical-polarization parameters is smaller than those corresponding values of light at vacuum state. Squeezing in HOPS is ascertained by referring the following basic inequalities (see Eq.(19)), derived in conjunction with Heisenberg uncertainty principle [26],

$$\langle (\Delta \hat{H}_0(t))^2 \rangle \text{ or } \langle (\Delta \hat{H}_2(t))^2 \rangle < |\langle \hat{H}_3 \rangle|,$$

$$\langle (\Delta \hat{H}_2(t))^2 \rangle \text{ or } \langle (\Delta \hat{H}_3(t))^2 \rangle < |1 + \langle \hat{H}_0 \rangle|,$$

$$\langle (\Delta \hat{H}_3(t))^2 \rangle \text{ or } \langle (\Delta \hat{H}_0(t))^2 \rangle < |\langle \hat{H}_2 \rangle|, \quad (23)$$

The quantum average values of Hidden optical-polarization operators, $\hat{H}_{0,1,2,3}$, and their variances at arbitrary time may be obtained by insertions of Eqs.(21) into Eq.(17). The results are given by

$$h_0(t) = \langle \hat{H}_0(t) \rangle = |\alpha_x|^2 (1 + |p_h|^2) C(4) - 2|\alpha_x|^2 |p_h| S(4) \sin\Delta_h + 2S^2(2), \quad (24)$$

$$h_1(t) = \langle \hat{H}_1(t) \rangle = |\alpha_x|^2 (|p_h|^2 - 1), \quad (25)$$

$$h_2(t) = \langle \hat{H}_2(t) \rangle = 2|\alpha_x|^2 |p_h| \cos\Delta_h, \quad (26)$$

$$h_3(t) = \langle \hat{H}_3(t) \rangle = 2|\alpha_x|^2 |p_h| C(4) \sin\Delta_h - \{1 + |\alpha_x|^2 (1 + |p_h|^2)\} S(4) \quad (27)$$

$$\langle (\Delta \hat{H}_0(t))^2 \rangle = |\alpha_x|^2 (1 + |p_h|^2) C(8) - 2|\alpha_x|^2 |p_h| S(8) \sin\Delta_h + S^2(4) \quad (28)$$

$$\langle (\Delta \hat{H}_1(t))^2 \rangle = |\alpha_x|^2 (1 + |p_h|^2), \quad (29)$$

$$\langle (\Delta \hat{H}_2(t))^2 \rangle = 1 + |\alpha_x|^2 (1 + |p_h|^2), \quad (30)$$

$$\langle (\Delta \hat{H}_3(t))^2 \rangle = |\alpha_x|^2 (1 + |p_h|^2) C(8) - 2|\alpha_x|^2 |p_h| S(8) \sin\Delta_h + S^2(4) - 1, \quad (31)$$

where $|p_h| = |\alpha_y|/|\alpha_x^*|$ and $\Delta_h = (\varphi_x + \varphi_y)$ denote the weird dependences on 'ratio of real amplitudes' and the 'sum of phases', the contrasting ingredients of HOPS. The condition for squeezing [39] may, associated with the measurement of $\hat{H}_2$, be accomplished by inequality $\langle (\Delta \hat{H}_2(t))^2 \rangle < | 1 + \langle \hat{H}_0 \rangle |$ in Eq.(23). The preceding inequality is satisfied if squeezing function, Sq bears the constraint, for a particular incident light,

$$\text{Sq}(kt, \Delta_h) > 1, \tag{32}$$

where Sq has explicit expression, exemplifying clear dependences on 'sum of phases', $\Delta_h$ and interaction time (kt), as

$$\text{Sq}(kt, \Delta_h) = \left|C(4) - \frac{2\,S(4)\sin\Delta_h}{|p_h|+|p_h|^{-1}(1+|\alpha_x|^{-2})}\right|. \tag{33}$$

Eq.(33) is numerically analyzed in Figs.1(a) & 1(b) which displays signatures of squeezing by departing from unit value. Squeezing in HOPS is observed to be a highly phase-sensitive phenomenon and, therefore, different onset-times for observation of squeezing exist depending upon equally likely values of 'sum of phases', $\Delta_h$.

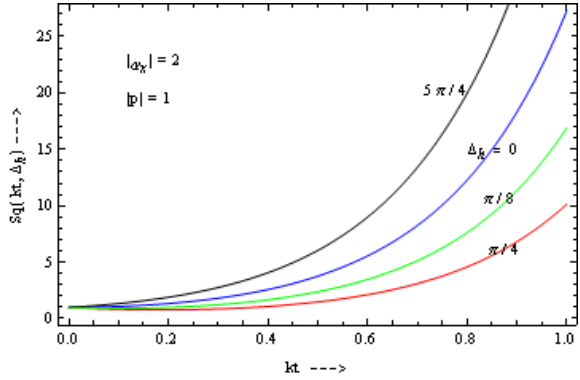
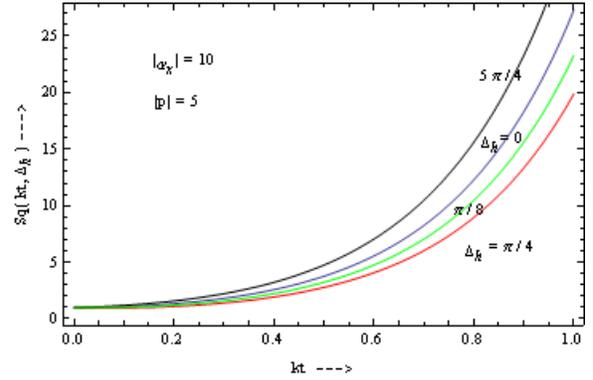

**Fig.1a** Squeezing function Sq showing critical dependences on $\Delta_h$, 'sum of phases' when the two modes consisting bi-modal coherent light is equally intense.

**Fig.1b** Squeezing function Sq showing critical dependences on $\Delta_h$, 'sum of phases' when the two modes consisting bi-modal coherent light has unequal intensities (here light in ($\hat{e}_y$, **k**) mode is twenty five times intense than that in ($\hat{e}_x$, **k**) mode).

## 6 Non-Classical Signature in HOPS

One may define, analogously to the Stokes degree of polarization, the degree of Hidden Optical-Polarization as

$$\mathcal{H}(t) = \frac{\left[h_1^2(t) + h_2^2(t) + h_3^2(t)\right]^{1/2}}{h_0(t)} \tag{34}$$

where $h_{0,1,2,3}(t)$ are given by Eqs.(24-27). Classically, $\mathcal{H}(t)$ should fall in the range of values, $0 \leq \mathcal{H}(t) \leq 1$ defining completely Hidden unpolarized state ($\mathcal{H} = 0$) and perfect Hidden optical-polarization ($\mathcal{H} = 1$). An alternative signature of non-classicality in HOPS may be setup if the degree of hidden optical-polarization, $\mathcal{H}(t)$ assumes values greater than unity, i. e., $\mathcal{H}(t) > 1$ at certain interaction time, t. After a simple yet straightforward calculations one arrives at time (t) satisfying,

$$t > t_0 = \frac{1}{2k} \tanh^{-1}\left[\frac{|p_h|\sin\Delta_h}{2\{|\alpha_x|^{-2}+(1+|p_h|^2)\}}\right], \tag{35}$$

and signifying the critical time ($t = t_0$) above which the generation of nonclassical Hidden optical-polarization state is attributed.

**Conclusion:-** HOPS is generated by utilizing bi-modal coherent light with orthogonally polarized collinearly propagating photons as pump field in Degenerate Parametric Amplification. Squeezing in Hidden Optical-Polarization is demonstrated by investigating the dynamic behavior of Hidden Optical-Polarization parameters. A Squeezing Function is recognized whose peculiar dependences on the interaction time for a particular equal and unequal intensities in the two modes of the light are demonstrated graphically. Present study of Non-Classicality, Dispersion (noise) and Squeezing in HOPS may be applied to characterize continuous-variable Hidden polarization entanglement which paves the way for application of HOPS in Quantum Cryptography.

**Acknowledgement**

The authors acknowledge gratefully for the critical yet inspirational discussions with Prof. H. Prakash and R. Prakash.